\documentclass[10pt]{iopart}

\pdfoutput=1

\usepackage{amssymb}
\usepackage{graphicx}
\usepackage{subfigure}
\usepackage{hyperref}

\eqnobysec

\begin{document}
\title[$r$-adic processes and fractal curves]
{Statistical properties of $r$-adic processes and their connections to
  families of popular fractal curves}
\author{J. R. Dorfman\dag, Thomas Gilbert\ddag}
\address{
  \dag Department of Physics and Institute for Physical Science and
  Technology, University of Maryland, College Park, MD 20742, USA\\
  \ddag Center for Nonlinear Phenomena and Complex Systems, Universit\'e
  Libre  de Bruxelles, C.~P.~231, Campus Plaine, B-1050 Brussels, Belgium}
\eads{\mailto{rdorfman@umd.edu}, \mailto{thomas.gilbert@ulb.ac.be}}

\begin{abstract}
  Results concerning the statists of $r$-adic processes and their
  fractal properties are reviewed. The connection between singular
  eigenstates of the statistical evolution of such processes and popular
  fractal curves is emphasized.
\end{abstract}

\noindent{\emph{In memory of our friend and colleague Shuichi Tasaki whose
    untimely death cut short a remarkable scientific career}} 
\vskip 0.5cm

Among the many important scientific achievements of Shuichi Tasaki were his
contributions to the statistical properties of piecewise linear maps and
the characterization of their eigenstates. In particular his work was key
to understanding the role played by singular measures in the statistical
evolution of chaotic maps and its connection to the mathematics of
fractals \cite{Antoniou:1992p11585, Tasaki:1993p4090}. This field, which
attracted much 
attention in the non-equilibrium statistical physics community in the
mid-1990's \cite{Gaspard:1998book, dorfman:1999book}, had been popular
among many Japanese mathematicians, in particular early in Tasaki's
scientific career \cite{Hata:1986p11282}. In this respect, it is
perhaps not surprising that one of Tasaki's favorite examples of fractals,
which helped understand how the non-equilibrium states of volume-preserving
strongly chaotic systems acquire fractal properties
\cite{Tasaki:1995p226}, was actually introduced by Teiji Takagi
\cite{Takagi:1903p1}, the founder of the school of modern mathematics in
Japan, who, more than a century ago, had proposed it as a simple example of
a continuous but nowhere differentiable function. 

It is the purpose of this article to review some of Tasaki's contributions
to this field, and draw observations which aim to underline the similarities
between the so-called hydrodynamic modes of diffusion of simple model systems
and some popular fractals, which include the von Koch and Levy curves.

\section{\label{sec.sngm}Singular measures of dyadic processes}

The statistical properties of coin tosses are full of mathematical wonders
and Shuichi Tasaki was well aware of it. Thus assume a fair coin tossing
game and a sequence $\{\omega_1,\dots,\omega_{k}\}$ of $k$ binary digits,
$\omega_i \in \{0,1\}$, $i = 1, \dots, k$, where say the 
symbol $0$ stands for heads and $1$ for tails. The probability measure
which assigns probability $1/2$ to every symbol irrespective of which of
heads or tails came at the previous toss is invariant under such a
process, which in particular is to say that every set of $k$ binary digits
has the same measure $1/2^k$. No surprise there. 

Consider however an arbitrary number
$p$, $0<p<1$, and let $\mu_p$ denote the probability measure which assigns
measure $p$ to heads and $1-p$ to tails. This measure is itself invariant
under fair coin tossings \cite{Eckmann:1985p234}. 

The reason is simply that the set $\{\omega_1,\dots,\omega_{k}\}$ is the
union of the two disjoint sets of length $k+1$, 
\begin{equation}
  \{0,\omega_1,\dots,\omega_k\} \cup
  \{1,\omega_1,\dots,\omega_k\}.
  \label{preimg-uok}
\end{equation}
Since $\mu_p(\{0, \omega_1, \dots, \omega_k\}) =
p\mu_p(\{\omega_1, \dots, \omega_k\})$ and
$\mu_p(\{1, \omega_1, \dots, \omega_k\}) = (1-p) \mu_p(\{\omega_1, \dots,
\omega_k\})$, we have that the probability measure of the left and right
hand sides of equation \eref{preimg-uok} is the same, irrespective of the
choice of the value $p$. We might say, in other words, that $\mu_p$ is as
good an invariant measure as the uniform measure $\mu_{1/2}$ is. 

\begin{figure}[htbp]
  \centering
  \subfigure[]{
    \includegraphics[width=.45\textwidth,angle=0]{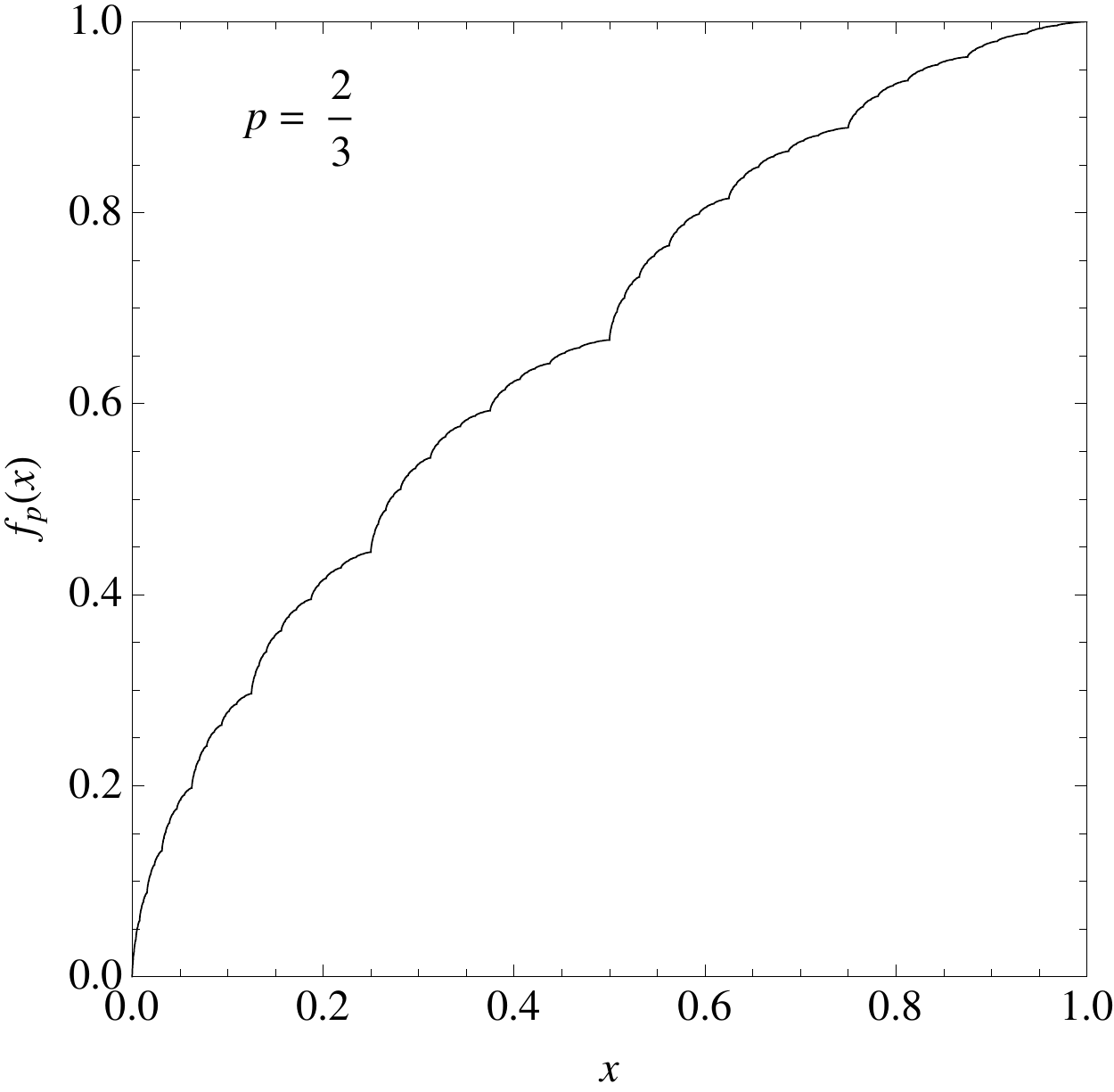}
    \label{fig.snglebsg1}
  }
  \subfigure[]{
    \includegraphics[width=.45\textwidth,angle=0]
    {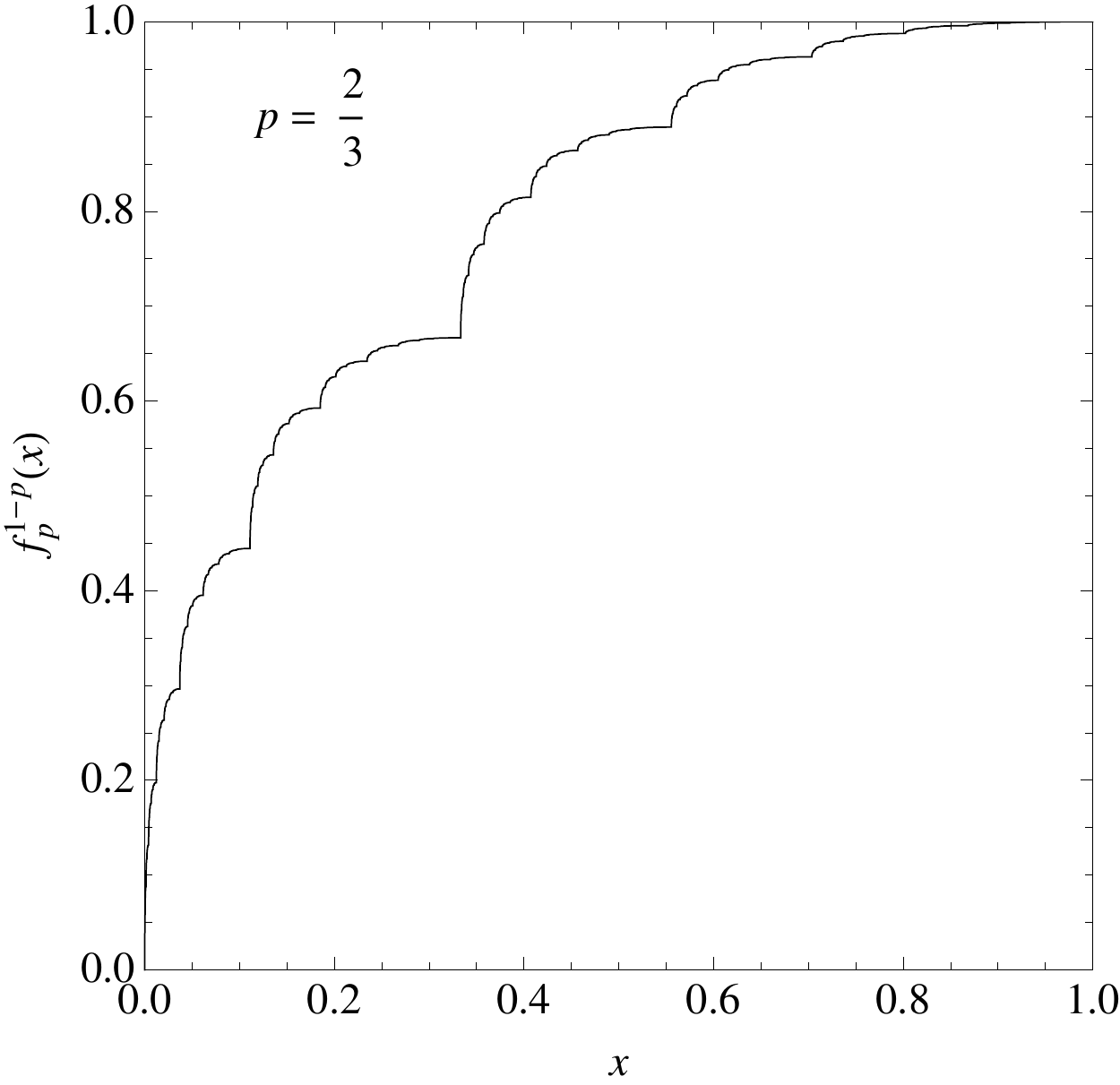}
    \label{fig.snglebsg2}
  }
  \caption{(a) Lebesgue's singular function $f_p$, equation \eref{snglebsg},
     and (b) $f_p^{(1-p)}$, equation \eref{snglebsgbias}, both with $p =
     2/3$. Here and below, unless otherwise stated, the curves are computed
     over $2^{14}$ points.}
  \label{fig.snglebsg}
\end{figure}

There is
however an essential difference here, which is that $\mu_{1/2}$ is the
natural invariant measure for the fair coin tossings. For $p\neq1/2$,
$\mu_p$ is in fact a singular measure, which can be thought of as a
singular Lebesgue function \cite{Tasaki:1998p339}. Using the isomorphism
between fair coin tossings and the angle-doubling map, $x\mapsto 2 x
(\mathrm{mod}\,1)$, it is a simple exercise to identify $\mu_p$ with the
measure lifted on the corresponding cylinder sets of the unit interval,
whose cumulant $\mu_p([0,x])$ is the unique function $f_p(x)$ satisfying 
\begin{equation}
  f_p(x) = 
  \left\{
  \begin{array}{l@{\quad}l}
    p f_p(2 x),& 0\leq x < 1/2,\\
    (1-p) f_p(2 x - 1) + p,& 1/2\leq x < 1.
  \end{array}
  \right.
  \label{snglebsg}
\end{equation}
Except for $p=1/2$, for which $f_{1/2}(x) = x$, $f_p$ is strictly
increasing and continuous, but has zero derivatives almost everywhere with
respect to the Lebesgue measure \cite{Bil65}. A specific example is shown
in \fref{fig.snglebsg1}.  

Notice however that $\mu_p$ is in fact the natural invariant measure of
the biased coin tossing which gives probability $p$ to heads and $1-p$ to 
tails. Its cumulant corresponds to a uniform measure, which
can be identified as the solution of the more general functional equation 
\begin{equation}
  f^{(q)}_p(x) = 
  \left\{
  \begin{array}{l@{\quad}l}
    p f^{(q)}_p(\frac{x}{q}),& 0\leq x < q,\\
    (1-p) f^{(q)}_p(\frac{x - q}{1-q}) + p,& q\leq x < 1.
  \end{array}
  \right.
  \label{snglebsgbias}
\end{equation}
Setting $q=p$ indeed yields $f^{(p)}_p(x) = x$. On the other hand, the
function $f_p^{(1-p)}$, illustrated in \fref{fig.snglebsg2}, is of the
Lebesgue singular type and arises as the cumulant of the natural invariant
measure of a dissipative baker map, projected along the contracting
direction \cite{Tasaki:1998p339}.

\section{\label{sec.cplx} Complex-valued measures of dyadic processes}

\begin{figure}[htbp]
  \centering
  \includegraphics[width=.5\textwidth,angle=0]{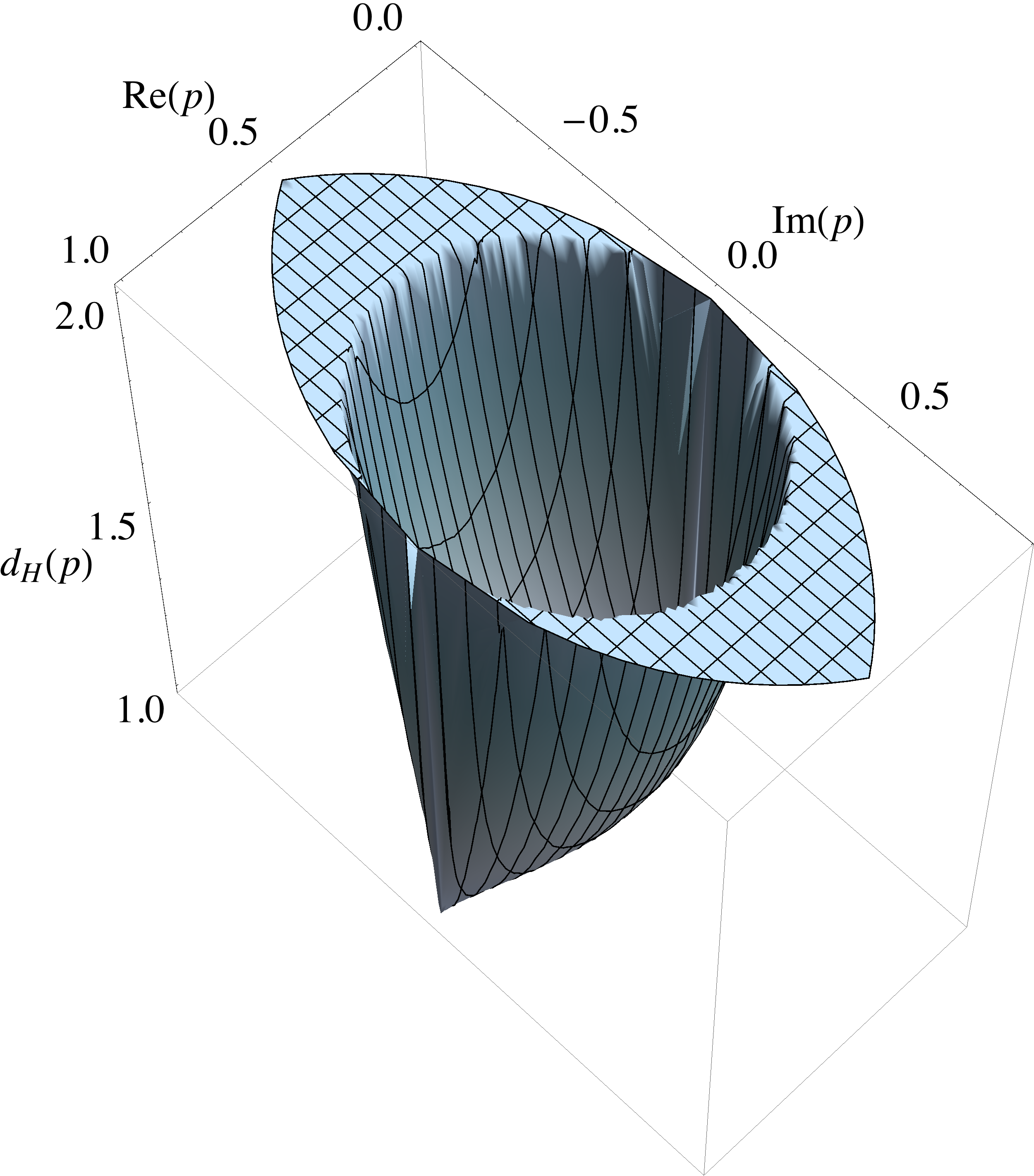}
  \caption{Hausdorff dimension \eref{DHsnglebsgtrans} of the curves $f_p$
    defined through equation \eref{snglebsg}.}
  \label{fig.snglebsgHD}
\end{figure}

Though equation \eref{snglebsg} is a functional equation characterizing a
probability measure, it is not in itself restricted to real values of the
parameter $p$. The uniqueness of its solutions is indeed warranted for
every complex parameter $p$ such that $|p|, |1-p|<1$ \cite{deRham:1957p1,
  deRham:1957p2}. In such cases the solutions of equation \eref{snglebsg}
are self-similar sets on the complex plane and it is a simple calculation
to show that their Hausdorff dimension is the solution
$d_\mathrm{H}$ of the following equation \cite{Tasaki:1994p11623},
\begin{equation}
  |p|^{d_\mathrm{H}} + |1-p|^{d_\mathrm{H}} = 1.
  \label{DHsnglebsgtrans}
\end{equation}
A graphical representation of its solution is shown in
\fref{fig.snglebsgHD}. 

Note in particular that the parameter values $p = 1/2(1+e^{\imath\varphi})$,
$0\leq\varphi<2\pi$, define a circle of complex parameters such that the
dimension \eref{DHsnglebsgtrans} is equal to $d_\mathrm{H} = 2$. Curves
with parameters whose values lie inside that circle have Hausdorff
dimension between $1$ and $2$. Curves with parameters which verify the
conditions $|p|<1$ and $|1-p| < 1$, but which lie outside that circle have
Hausdorff dimension $2$ \cite{deRham:1957p2}

As shown by Tasaki and collaborators \cite{Tasaki:1993p4090}, the functions
$f_p$ with complex parameter  
\begin{equation}
  p(k) = \frac{e^{\imath k}}{2 \cos k} = \frac{1}{2}\left(1 + \imath \tan k
    \right),
  \label{snglebsgpq}
\end{equation}
with $k$ a real number such that $|k|<\pi/3$, occur as representations of
the left eigenvectors of the Frobenius-Perron operator of an expanding
piecewise linear map of the real line related to the angle-doubling
map. 

The same curves also occur as special cases of hydrodynamic modes of
diffusion, associated with multi-baker maps on a ring
\cite{Gilbert:2001p356}, where $k$ is the associated wavenumber. The
dynamics is defined according to 
\begin{equation}
  (n,x,y) \mapsto
  \left\{
    \begin{array}{l@{\quad}l}
      (n - 1, 2x, \frac{y}{2}),&0\leq x < 1/2,\\
      (n + 1, 2x -1, \frac{y + 1}{2}),&1/2\leq x < 1,
    \end{array}
  \right.
  \label{mbaker}
\end{equation}
where $n = 0,\dots, N - 1$ and $n\pm1$ are understood to be taken modulo
$N$. Given an initial distribution, the relaxation of statistical ensembles
to the uniform equilibrium measure takes place exponentially fast at rate
$\sim N^{-2}$ and is
best characterized in terms of the cumulant measure $\mu_t(n, [0,1],[0,y])$
of phase points $(n, \tilde x, \tilde y)$, with $0\leq \tilde x < 1$ and $0
\leq \tilde y < y$, after $t$ iterations:   
\begin{equation}
  \mu_t(n, [0,1],[0,y]) = \sum_k a_k \cos^t(k) F_k(y) e^{\imath k n},
  \label{hydromodes}  
\end{equation}
where $k = 2\pi m/N$, $m = 0,\dots, N-1$ are the wavenumbers, the
coefficients $a_k$ are set by the initial distribution, and $F_k$ are
solutions of the system of equations \eref{snglebsg} with the choice of
parameter \eref{snglebsgpq}, $F_k(x) \equiv f_{p(k)}(x)$, \emph{viz.}
\begin{equation}
  F_k(x) = 
  \left\{
  \begin{array}{l@{\quad}l}
    \frac{e^{\imath k}}{2 \cos k} F_k(2 x),& 0\leq x < 1/2,\\    
    \frac{e^{-\imath k}}{2 \cos k} F_k(2 x - 1) + 
    \frac{e^{\imath k}}{2 \cos k},& 1/2\leq x < 1.
  \end{array}
  \right.
  \label{snglebsgFk}
\end{equation}
The dimension \eref{DHsnglebsgtrans} of these curves can be computed
explicitly \cite{Tasaki:1993p4090}:
\begin{equation}
  d_\mathrm{H}(k) = \frac{\log 2}{\log (2|\cos k|)}.
  \label{DHsnglebsg}
\end{equation}

\begin{figure}[htbp]
  \centering
  \subfigure[Fractal curve with dimension $\log4/\log3$]{
    \includegraphics[width=.45\textwidth,angle=0]{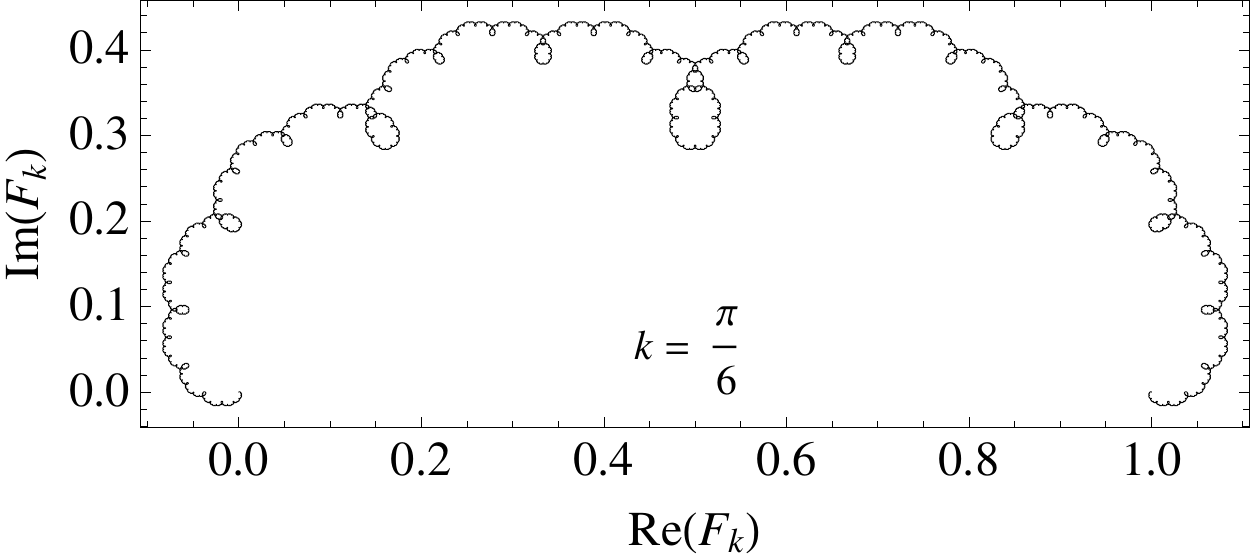}
    \label{fig.levy1}
  }
  \subfigure[Levy dragon with dimension $2$]{
    \includegraphics[width=.45\textwidth,angle=0]{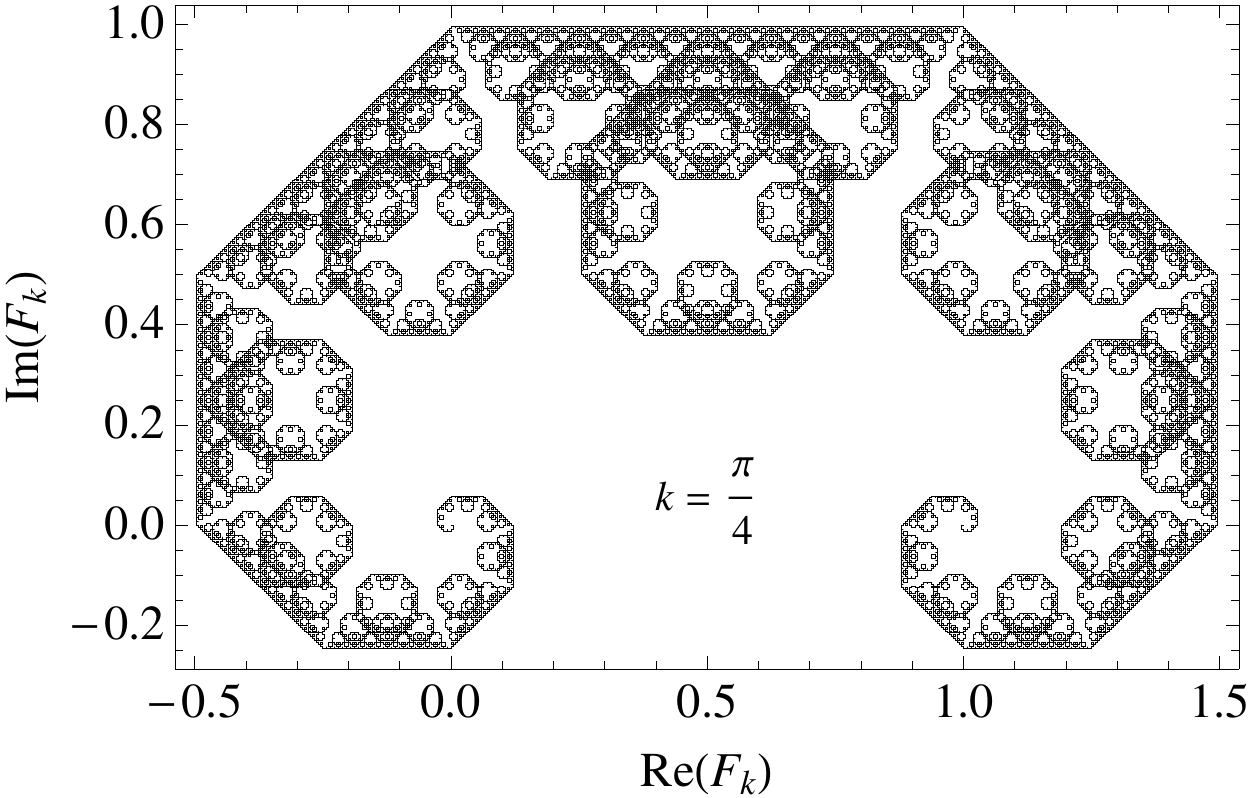}
    \label{fig.levy2}
  }
  \caption{Specific examples of hydrodynamic modes for the dyadic process
    defined by the functional equations \eref{snglebsgFk}.}
  \label{fig.levy}
\end{figure}

In particular, for $k = \pi/4$, we obtain a curve of Hausdorff 
dimension $d_\mathrm{H}(\pi/4) = 2$ which covers positive areas of the plane
and corresponds to the Levy dragon \cite{Levy:1938p1}. Another remarkable
value of the parameter is $k = \pi/6$, for which we have the dimension
$d_\mathrm{H}(\pi/6) = \log4/\log3$. These two curves are displayed in
\fref{fig.levy}.

\begin{figure}[htbp]
  \centering
  \subfigure[]{
    \includegraphics[width=.45\textwidth,angle=0]{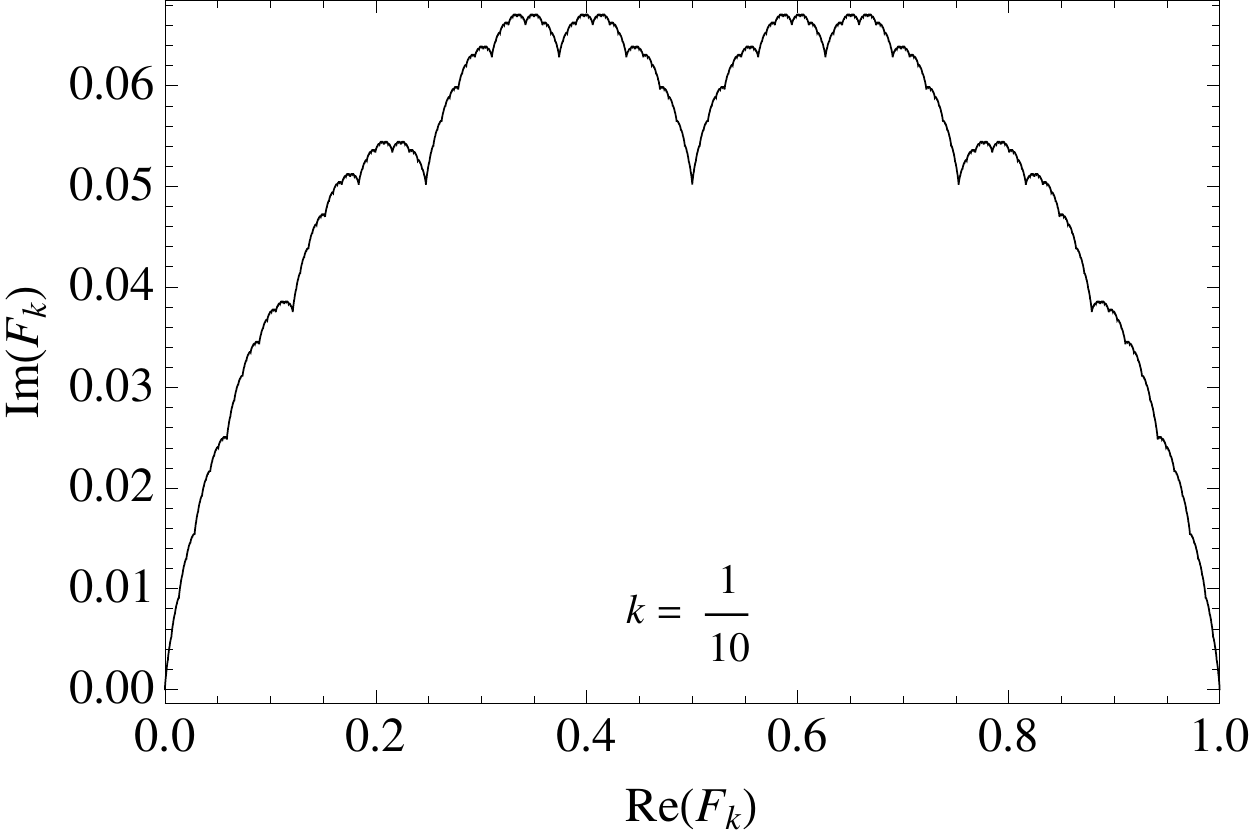}
  }
  \subfigure[]{
    \includegraphics[width=.45\textwidth,angle=0]{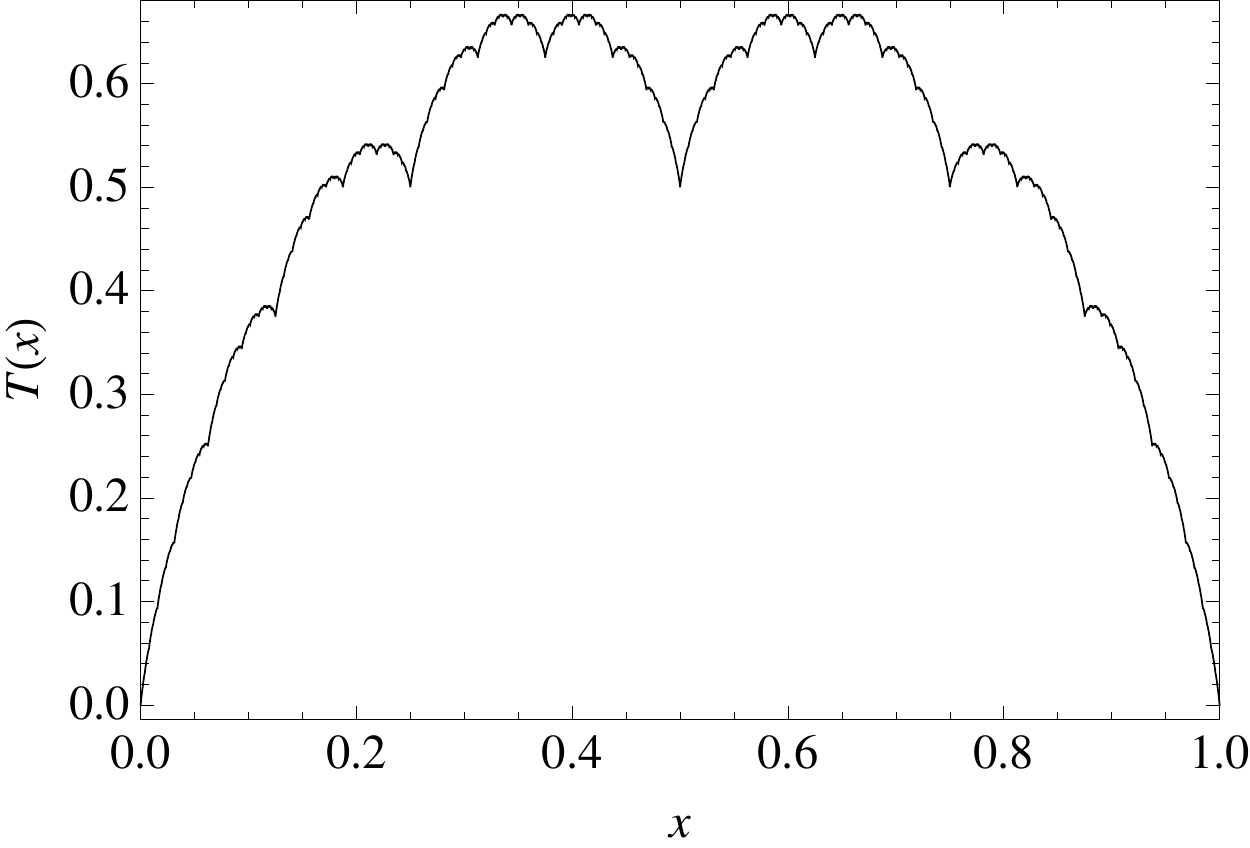}
  }
  \caption{The graph of the function $F_k$, equation
    \eref{snglebsgFk}, with $k =  1/10$ (a), approaches that of the
    Takagi function \eref{takagi} (b).}
  \label{fig.takagi}
\end{figure}

In the context of hydrodynamic modes of diffusion \cite{Gaspard:2001p333},
the complex curves defined by equation \eref{snglebsg} acquire a 
clear physical meaning , which comes about when the parameter $k$ in
\eref{snglebsgpq} is small: $|k|\ll1$. The Hausdorff dimension
\eref{DHsnglebsg} then becomes \cite{Gilbert:2001p356} 
\begin{equation}
  d_\mathrm{H}(k) = 1 + \frac{k^2}{2\log 2} + \mathcal{O}(k^4),
  \label{DHsnglebsglim}
\end{equation}
which provides a relation between the Hausdorff dimension of the
hydrodynamic modes, the diffusion coefficient of the process and the
positive Lyapunov exponent of the underlying dynamics
\eref{mbaker}. Furthermore, 
considering the functional equation \eref{snglebsgFk}, we obtain the 
equivalent of a gradient expansion for $F_k$ \cite{Gilbert:2000p355},

\begin{equation}
  F_{k}(x) = x + \imath k T(x) + \mathcal{O}(k^2),
  \label{limitsnglebsgFk}
\end{equation}
where $T(x)$ is the Takagi function \cite{Takagi:1903p1}, which can be
defined through the functional equation:
\begin{equation}
  T(x) = 
  \left\{
  \begin{array}{l@{\quad}l}
    x + \frac{1}{2} T(2 x),& 0\leq x < 1/2,\\
    1 - x + \frac{1}{2} T(2 x - 1),& 1/2\leq x < 1.
  \end{array}
  \right.
  \label{takagi}
\end{equation}
See \fref{fig.takagi}.

The result \eref{limitsnglebsgFk} should be compared with that of Hata and
Yamaguti \cite{Yamaguti:1983p1, Hata:1984p11424}, who
recover the Takagi function as the derivative of the curves $f_p$,
equation \eref{snglebsg}, with respect to their parameter, $T(x) = \partial
f_p(x)/\partial p|_{p = 1/2}$. 

\section{\label{sec.}Levy, von Koch and Heighway curves}

The Levy dragon shown in \fref{fig.levy2} is but one example in a long list
of popular fractal curves which are most often defined recursively through
iterated function systems \cite{Edgar:2008p11496} or using $L$-systems
\cite{Prusinkiewicz:1985p11492}. The von Koch curve \cite{vonKoch:1904p1}
and the Heighway dragon, popularized in Martin Gardner's
\emph{Mathematical games} \cite{Gardner:1989}, are
two other similar examples. In 
analogy with equation \eref{snglebsg}, it is simple to identify 
linear contractions which generate these sets \cite{Hata:1985p11486,
  Yamaguti:1997}.

\begin{figure}[htbp]
  \centering
  \subfigure[]{
    \includegraphics[width=\textwidth,angle=0]{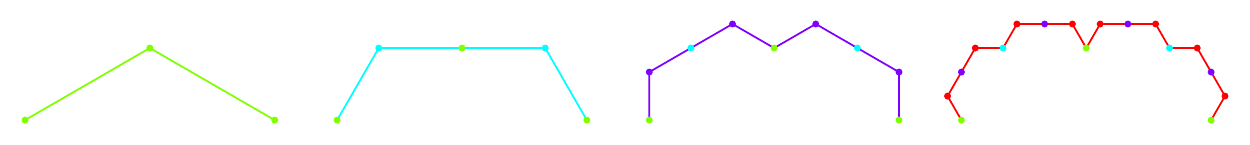}
    \label{fig.itlevy}
  }  
  \subfigure[]{
    \includegraphics[width=\textwidth,angle=0]{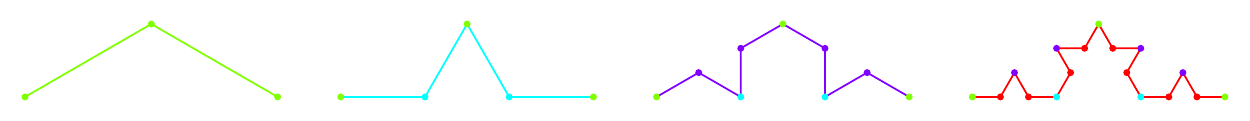}
    \label{fig.itkoch}
  }
  \caption{Comparison between the first few steps of the iterative
    constructions of (a) the Levy curve and (b) the von Koch curve.}
  \label{fig.iter}
\end{figure}

The solutions of equation \eref{snglebsg} with complex parameters produce a
family of curves somewhat similar to the Levy dragon. Looking at the
iterative construction of these solutions, we start from the initial line
segment joining $F_k(0) = 0$ and $F_k(1) = 1$ and obtain $F_k(1/2) =
e^{\imath k}/(2\cos k)$. At the next iteration, we add the two points
$F_k(1/4)$ and $F_k(3/4)$. The result on the complex plane is a collection
of two isosceles triangles whose long edges correspond to the two smaller
edges of the triangle formed by the three initial points $F_k(0)$,
$F_k(1/2)$ and $F_k(1)$. The next iteration produces four smaller isosceles
triangles which are stacked upon the small edges of the two existing
ones. See \fref{fig.itlevy}.

A similar construction, which consists of stacking up the triangles
alternatively upwards and downwards from one iteration to the next, as
shown in \fref{fig.itkoch}, produces curves similar to the von Koch
curve. They are most easily obtained as the self-similar sets associated
with a linear contraction similar to equation \eref{snglebsg}, but that
further involves complex conjugation:
\begin{equation}
  g_p(x) = 
  \left\{
  \begin{array}{l@{\quad}l}
    p g_p^*(2 x),& 0\leq x < 1/2,\\
    (1-p) g_p^*(2 x - 1) + p,& 1/2\leq x < 1,
  \end{array}
  \right.
  \label{snglebsg2}
\end{equation}
where $^*$ denotes the complex conjugation. Equations \eref{snglebsg} and
\eref{snglebsg2} share the same real solutions but have different sets of
solutions for complex parameter values. Nonetheless they share the 
same Hausdorff dimension \eref{DHsnglebsgtrans}. As pointed out by de Rham 
\cite{deRham:1957p2}, the choice $p = 1/2(1 + \imath /\sqrt3)$ produces the
von Koch curve.

\begin{figure}[htbp]
  \centering
  \subfigure[]{
    \includegraphics[width=.45\textwidth,angle=0]{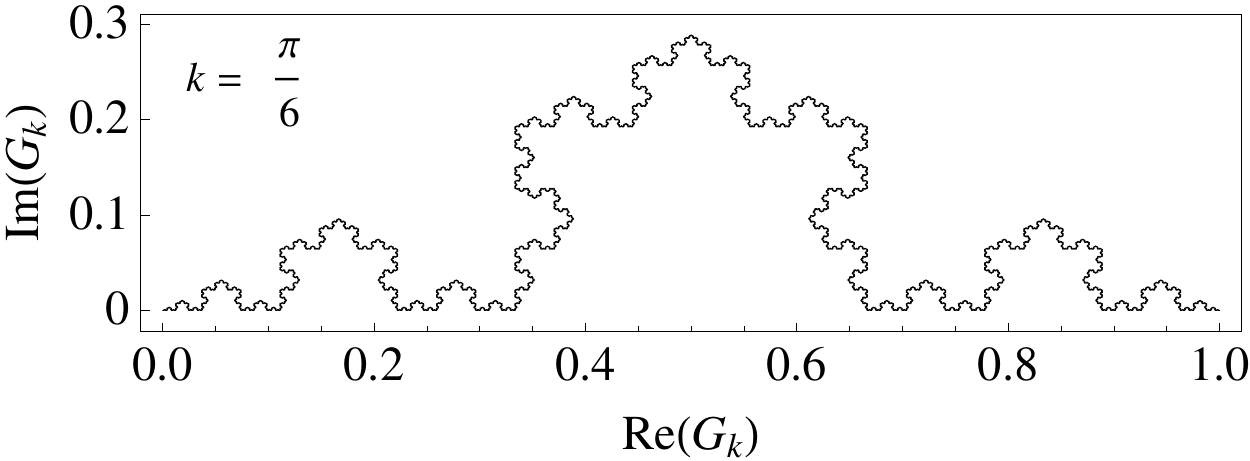}
  }
  \subfigure[]{
    \includegraphics[width=.45\textwidth,angle=0]{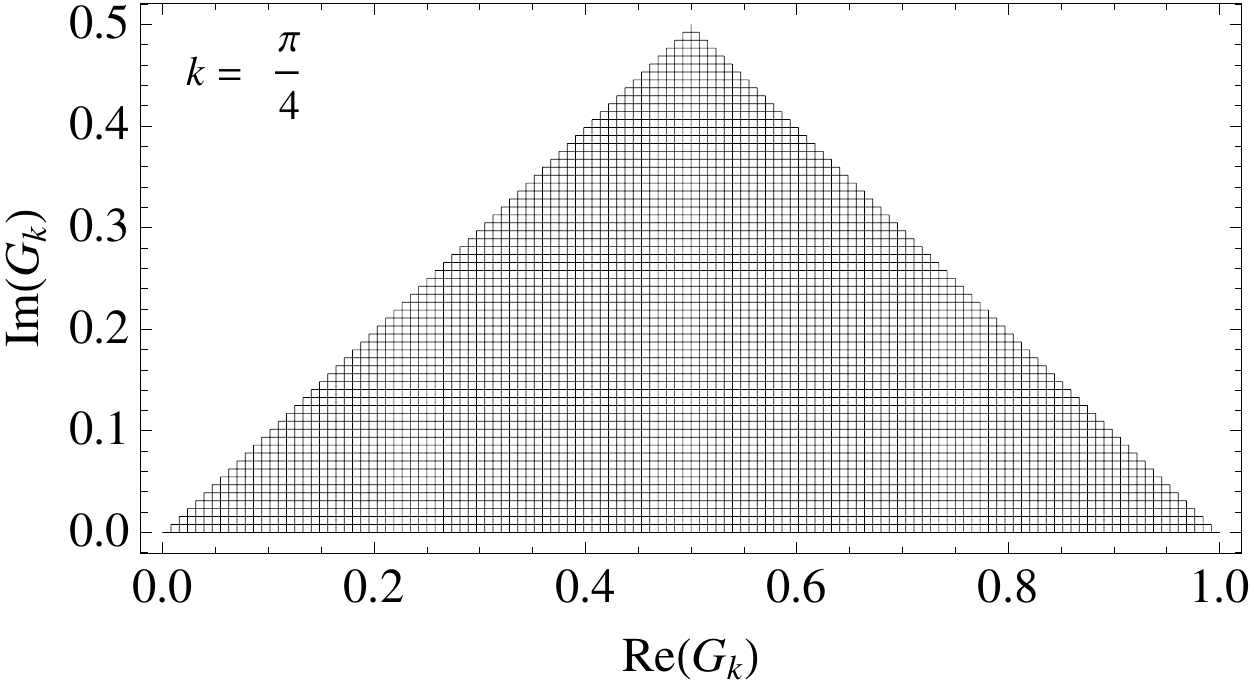}
  }
  \caption{Specific examples of solutions of the functional equation
    \eref{snglebsgGk} yield (a) the von Koch curve and (b) a Peano-type
    curve.}
  \label{fig.koch}
\end{figure}

Taking the parameter $p$ as defined by equation \eref{snglebsgpq}, we write
$G_k(x) \equiv g_{p(k)}(x)$, for which we have the functional
equation: 
\begin{equation}
  G_k(x) = 
  \left\{
  \begin{array}{l@{\quad}l}
    \frac{e^{\imath k}}{2 \cos k} G_k^*(2 x),& 0\leq x < 1/2,\\
    \frac{e^{-\imath k}}{2 \cos k} G_k^*(2 x - 1) + 
    \frac{e^{\imath k}}{2 \cos k)},& 1/2\leq x < 1.
  \end{array}
  \right.
  \label{snglebsgGk}
\end{equation}
As illustrated in \fref{fig.koch}, the choice of parameter $k = \pi/6$
corresponds to the von Koch curve. For $k = \pi/4$ we obtain a Peano-like
curve.

As such, the functional equation \eref{snglebsgGk} is different from
equation \eref{snglebsgFk} so its solutions do not correspond to
hydrodynamic modes of the dyadic multi-baker map \eref{mbaker}. It is
however straightforward to see that they are the hydrodynamic modes
associated with the four-adic map, 
\begin{equation}
  (n,x,y) \mapsto
  \left\{
    \begin{array}{l@{\quad}l}
      (n, 4x, \frac{y}{4}),&0\leq x < 1/4,\\
      (n - 1, 4x - 1, \frac{y+1}{4}),&1/4\leq x < 1/2,\\
      (n + 1, 4x - 2, \frac{y+2}{4}),&1/2\leq x < 3/4,\\
      (n, 4x -3, \frac{y + 3}{4}),&3/4\leq x < 1,
    \end{array}
  \right.
  \label{m4baker}
\end{equation}
with wavenumbers halved. Note that the iterative construction of the von
Koch curve using a four-adic process amounts to skipping every odd step in
the dyadic-based iterative construction shown in \fref{fig.itkoch}. The
former is often preferred over the latter in the literature, see for
instance \cite{Benenson:2002}, even though the dyadic representation based
on equation \eref{snglebsgGk} is indeed the most compact. 

\begin{figure}[htbp]
  \centering
  \subfigure[]{
    \includegraphics[width=\textwidth,angle=0]{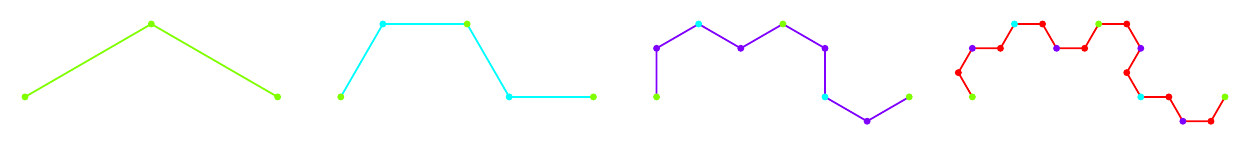}
    \label{fig.itpm}
  }
  \subfigure[]{
    \includegraphics[width=\textwidth,angle=0]{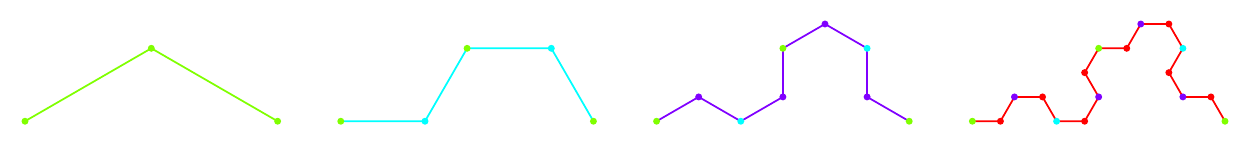}
    \label{fig.itmp}
  }
  \caption{First few steps of the iterative constructions of asymmetric
    curves similar to the the Levy and von Koch constructions of
    \fref{fig.iter}.} 
  \label{fig.iter2}
\end{figure}

The Heighway dragon and related curves can also be obtained through
functional equations similar to equation \eref{snglebsg}. They are based
upon iterative constructions which combine both upwards and downwards
triangles, as shown in \fref{fig.iter2}. 

The Heighway dragon is in fact a particular case of a hydrodynamic mode of
a dyadic map similar to the multibaker map \eref{mbaker}, where the
angle-doubling map is replaced by the tent map $x\mapsto 2x$ if $0\leq x
<1/2$ and $2(1-x)$ if $1/2 \leq x < 1$, namely
\begin{equation}
  (n,x,y) \mapsto
  \left\{
    \begin{array}{l@{\quad}l}
      (n - 1, 1 - 2x, \frac{y}{2}),&0\leq x < 1/2,\\
      (n + 1, 2x -1, 1 - \frac{y}{2}),& 1/2\leq x < 1.
    \end{array}
  \right.
  \label{mtent}
\end{equation}
Note that the tent map here appears upside down along the expanding
direction. This choice of combination of maps along the $x$ and $y$
coordinates is taken so the map has the property of being time reversal
symmetric under the induction $(n,x,y)\mapsto(n,1-y,1-x)$, which is also
the case of the multi-baker map \eref{mbaker}. Another such map, very
similar to equation \eref{mtent}, is  
  \begin{equation}
    (n,x,y) \mapsto
    \left\{
      \begin{array}{l@{\quad}l}
        (n - 1, 2x, \frac{1-y}{2}),&0\leq x < 1/2,\\
        (n + 1, 2(1-x), \frac{y+1}{2}),&1/2\leq x < 1.
      \end{array}
    \right.
    \label{mtent2}
\end{equation}
Here the usual tent map acts along the expanding direction. For the sake of
our argument however, we prefer using \eref{mtent} since the tent map
appears along the contracting direction under the time-evolution of
phase-space densities. 

\begin{figure}[htbp]
  \centering
  \subfigure[]{
    \includegraphics[width=.45\textwidth,angle=0]{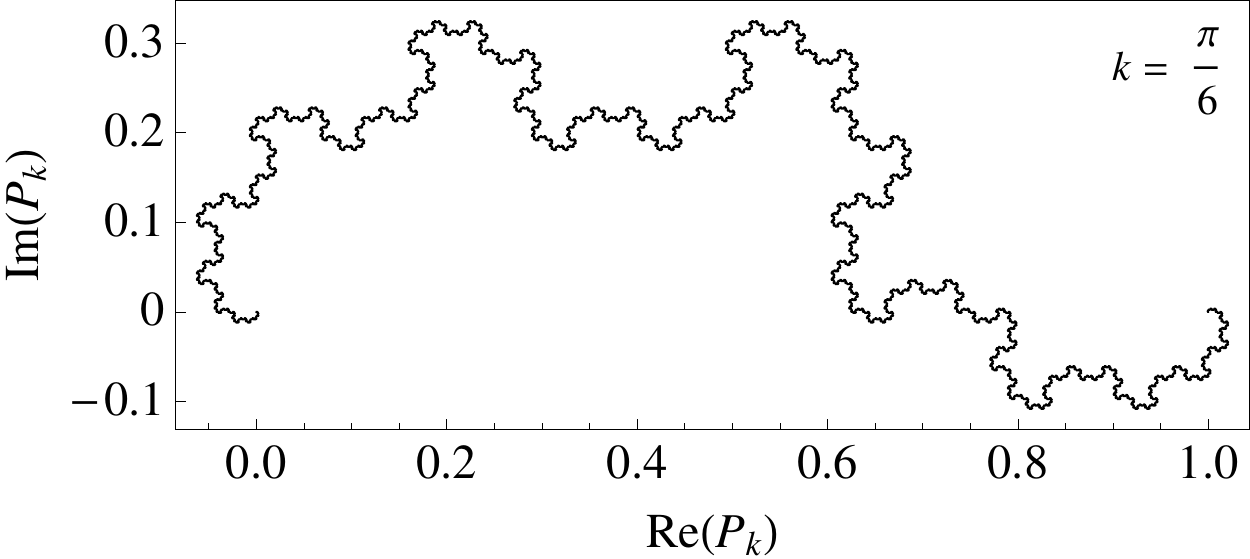}
  }
  \subfigure[]{
    \includegraphics[width=.45\textwidth,angle=0]{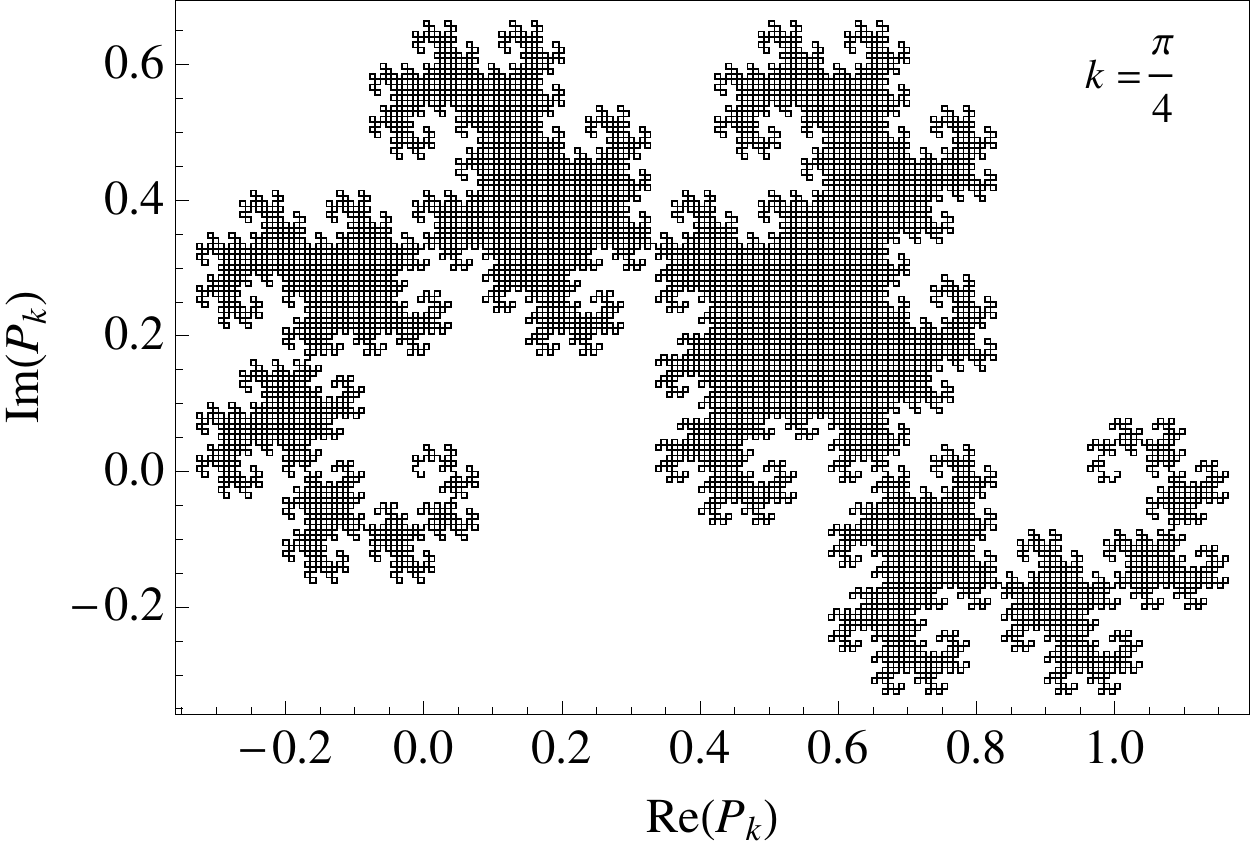}
  }
  \caption{Examples of different solutions of equation \eref{snglebsgPk}
    with (a) $k = \pi/6$ and (b) $k = \pi/4$. The latter is knwon as the
    Heighway dragon}
  \label{fig.heighway}
\end{figure}

Using, for the evolution of statistical ensembles under \eref{mtent}, an
expansion similar to equation \eref{hydromodes}, we identify the modes
\begin{equation}
  P_k(x) = 
  \left\{
  \begin{array}{l@{\quad}l}
    \frac{e^{\imath k}}{2 \cos k} P_k (2 x),& 0\leq x < 1/2,\\
    1 - \frac{e^{-\imath k}}{2 \cos k} P_k(2 (1-x)),& 1/2\leq x < 1.
  \end{array}
  \right.
  \label{snglebsgPk}
\end{equation}
The value $k = \pi/4$ yields the Heighway dragon, as shown in
\fref{fig.heighway}. The map \eref{mtent2} produces modes which can be
obtained from equation \eref{snglebsgPk} by a simple symmetry.

\begin{figure}[htbp]
  \centering 
  \subfigure[]{
    \includegraphics[width=.45\textwidth,angle=0]{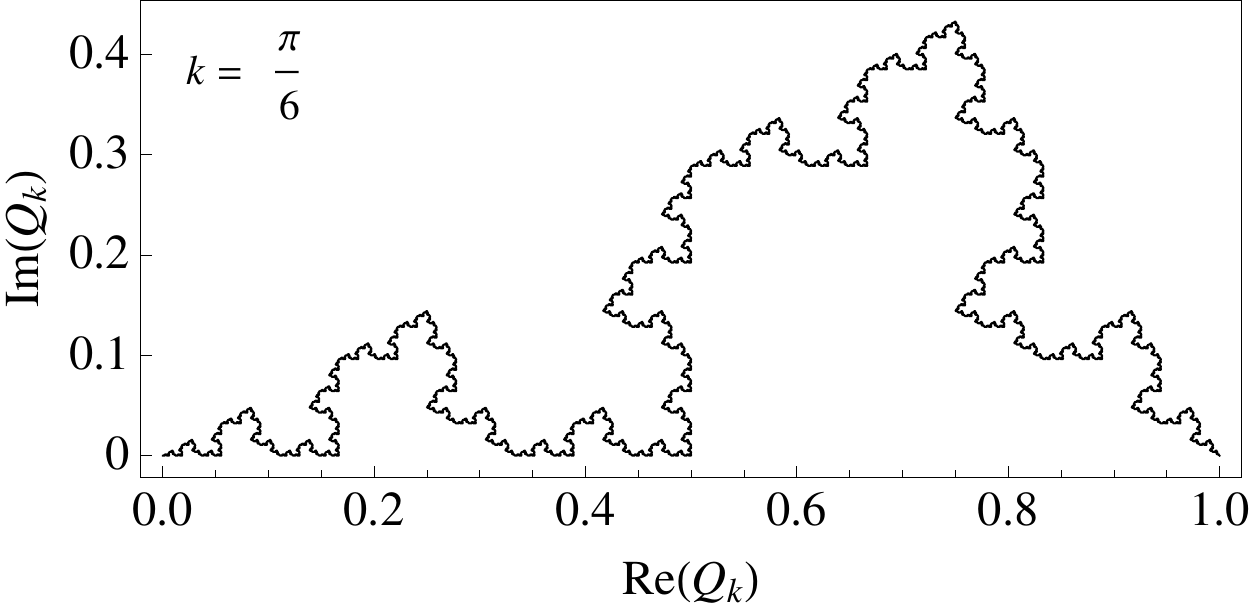}
  }
  \subfigure[]{
    \includegraphics[width=.45\textwidth,angle=0]{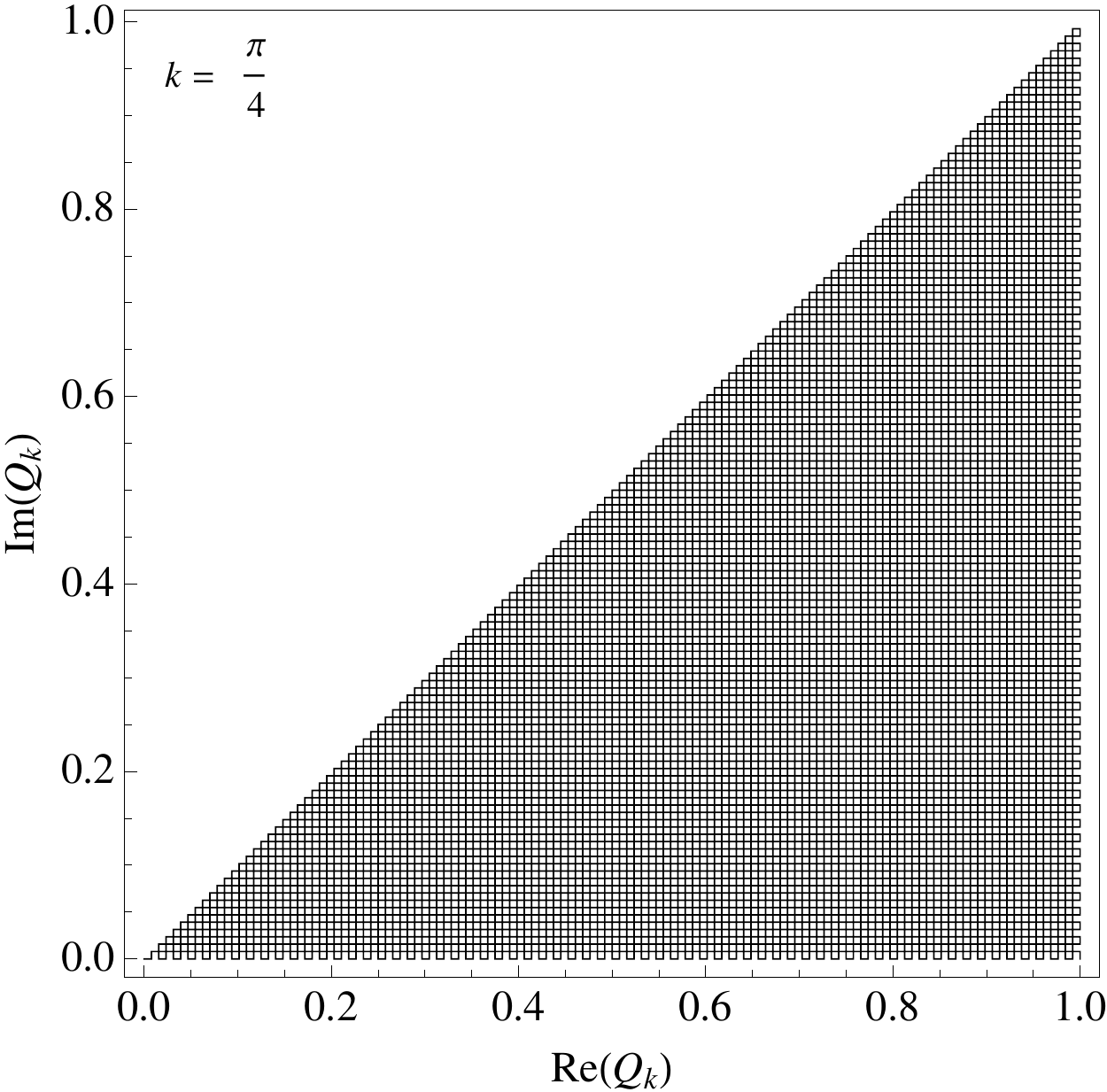}
  }
  \caption{Two different solutions of equation \eref{snglebsgQk} with (a)
    $k = \pi/6$ and (b) $k = \pi/4$, which is another example of Peano-type
    curve.} 
  \label{fig.conjheighway}
\end{figure}

Finally, by analogy with equation \eref{snglebsgGk}, we obtain another set
of associated fractals by complex conjugating the function on the
right-hand side of the functional equation,
\begin{equation}
  Q_k(x) = 
  \left\{
    \begin{array}{l@{\quad}l}
      \frac{e^{\imath k}}{2 \cos k} Q^*_k (2 x),& 0\leq x < 1/2,\\
      1 - \frac{e^{-\imath k}}{2 \cos k} Q^*_k(2 (1-x)),& 1/2\leq x < 1.
    \end{array}
  \right.
  \label{snglebsgQk}
\end{equation}
Examples of solutions are shown in \fref{fig.conjheighway}. Here again
these curves can be identified as the hydrodynamic modes of a four-adic
map. 

\section{\label{sec.radic}Complex measures of $r$-adic processes}

\begin{figure}[htbp]
  \centering
  \subfigure[]{
    \includegraphics[width=.95\textwidth,angle=0]{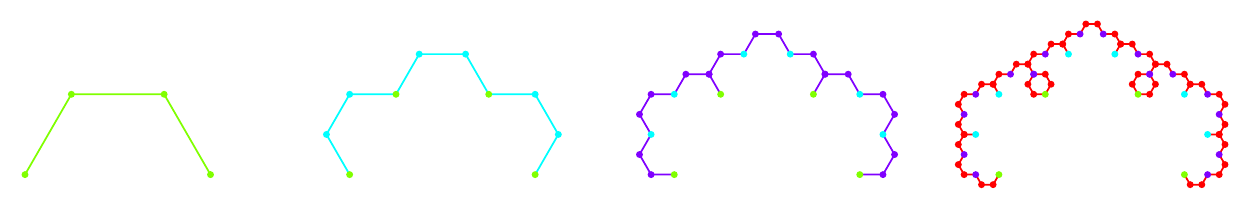}
    \label{fig.itggd}
  }
  \subfigure[]{
    \includegraphics[width=.95\textwidth,angle=0]{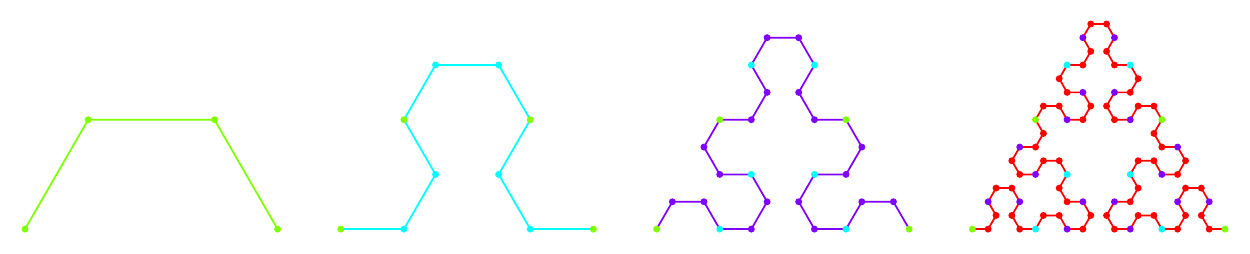}
    \label{fig.itsierp}
  }
  \subfigure[]{
    \includegraphics[width=.95\textwidth,angle=0]{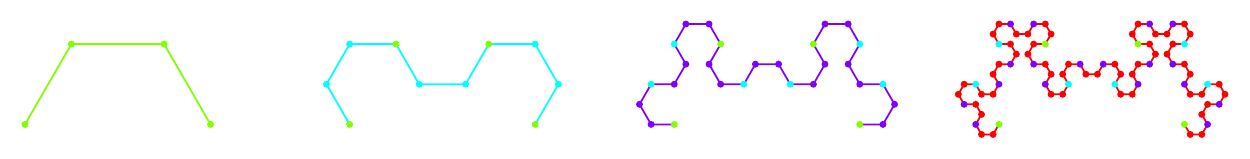}
    \label{fig.itpmp}
  }
  \caption{These three examples of iterative constructions of fractal curves
    are based upon triadic processes which can be conveniently written under
    the form of functional equation systems \eref{snglebsgUk}-\eref{snglebsgWk}.}
  \label{fig.iter3}
\end{figure}

\begin{figure}[htbp]
  \centering
  \subfigure{
    \includegraphics[width=.31\textwidth,angle=0]{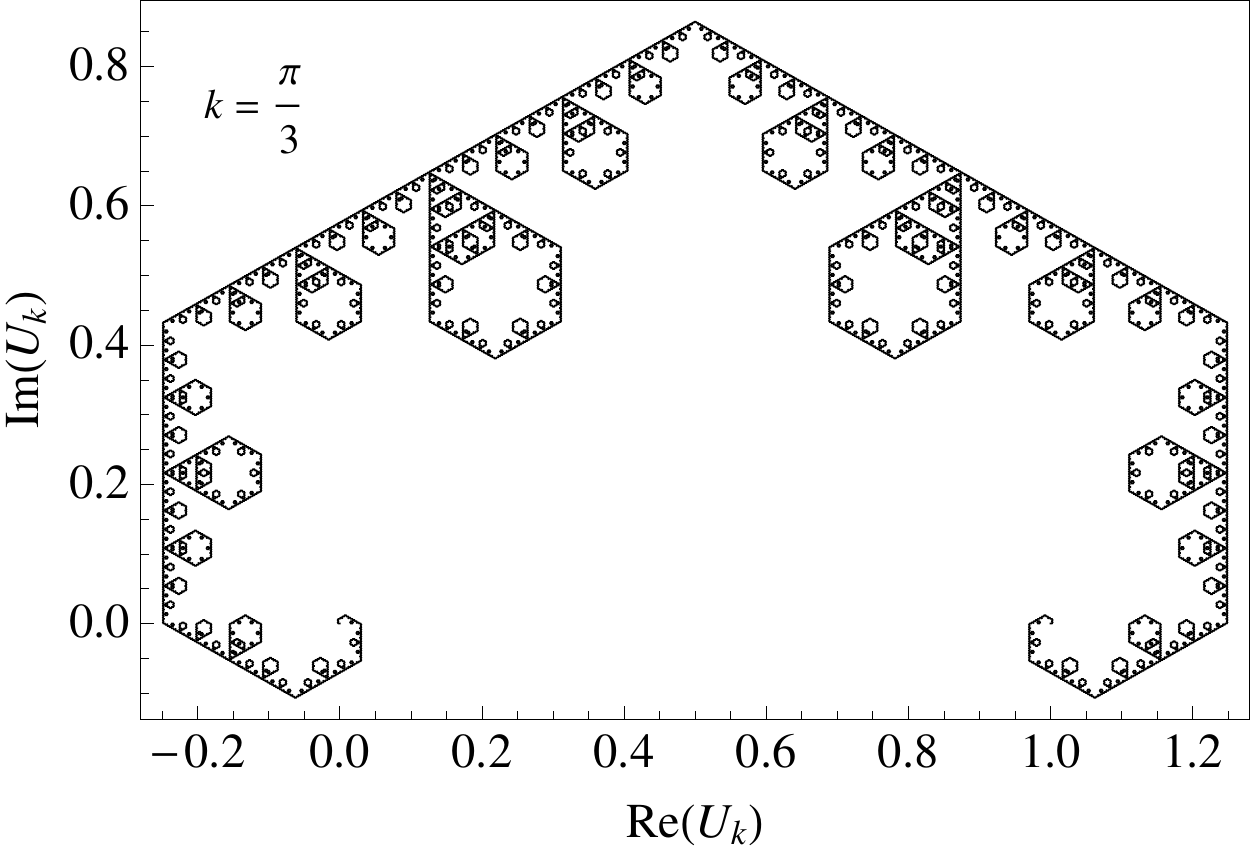}
  }
  \subfigure{
    \includegraphics[width=.31\textwidth,angle=0]{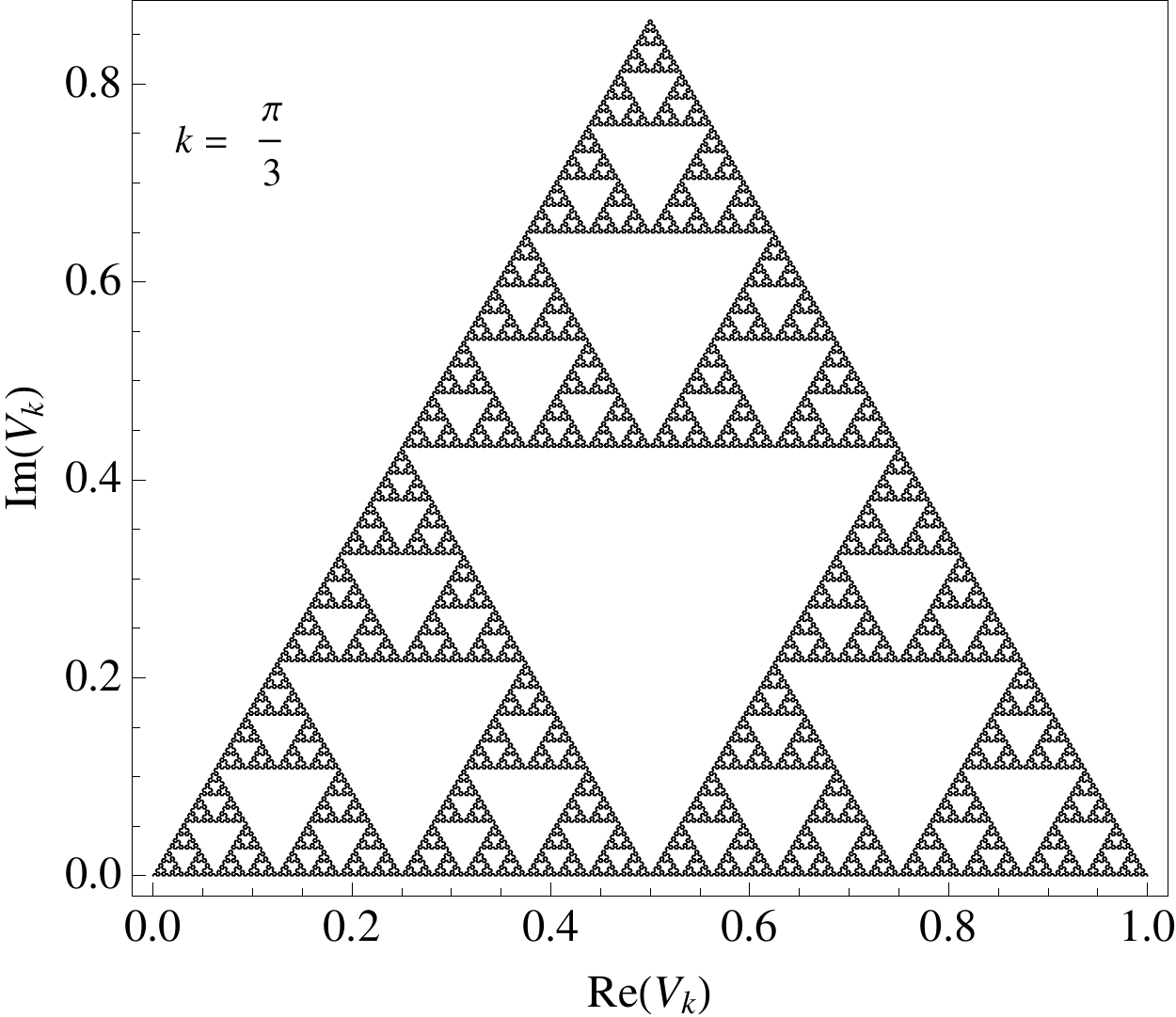}
  }
  \subfigure{
    \includegraphics[width=.31\textwidth,angle=0]{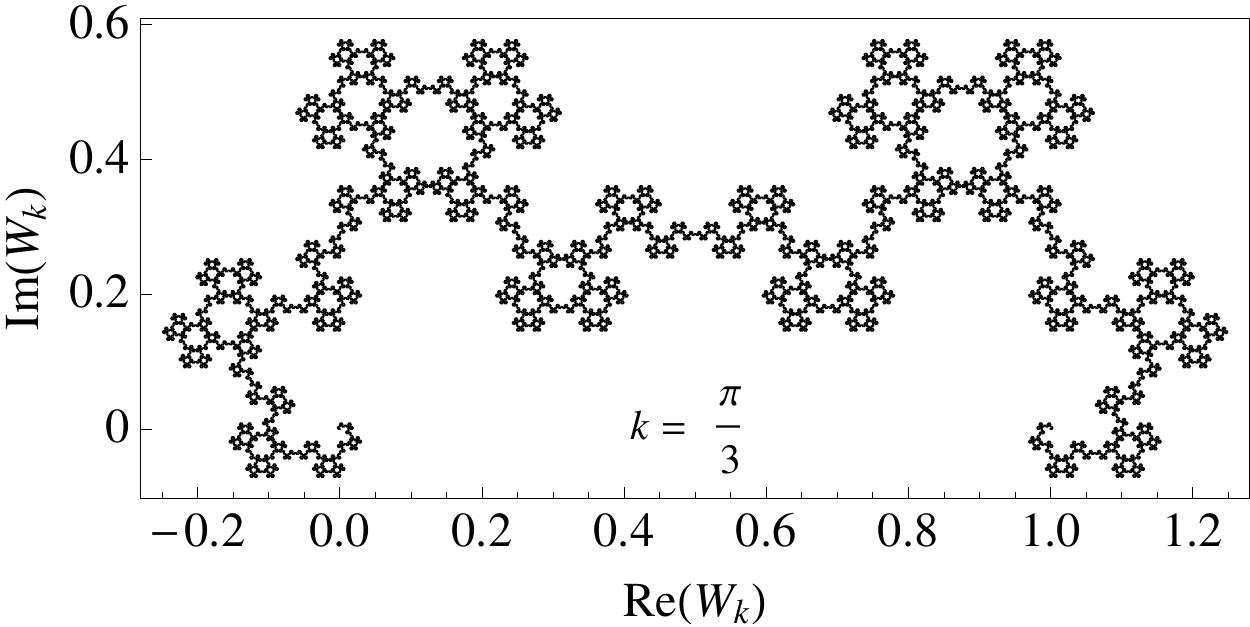}
  }
  \caption{Examples of different solutions of equations (a)
    \eref{snglebsgUk}, (b) \eref{snglebsgVk}, and (c) \eref{snglebsgWk},
    with $k = \pi/3$, here computed over $3^{9}$ points.} 
  \label{fig.sierp}
\end{figure}

The above considerations are not limited to dyadic processes. Consider for
instance the iterated systems shown in \fref{fig.iter3}
The functional equations whose solutions reproduce these  familiar curves
are based upon triadic processes:
\begin{eqnarray}
  U_k(x) = 
  \left\{
    \begin{array}{l@{\quad}l}
      \frac{e^{\imath k}}{1 + 2 \cos k} U_k (3 x),& 0\leq x < 1/3,\\
      \frac{e^{\imath k}}{1 + 2 \cos k} + 
      \frac{1}{1 + 2 \cos k} U_k (3x - 1),& 1/3\leq x <
      2/3,\\
      \frac{e^{\imath k} + 1}{1 + 2 \cos k} + 
      \frac{e^{-\imath k}}{1 + 2 \cos k} U_k(3x - 2),& 2/3\leq x <
      1,
    \end{array}
  \right.
  \label{snglebsgUk}
  \\
  V_k(x) = 
  \left\{
    \begin{array}{l@{\quad}l}
      \frac{e^{\imath k}}{1 + 2 \cos k} V^*_k (3 x),& 0\leq x < 1/3,\\
      \frac{e^{\imath k}}{1 + 2 \cos k} + 
      \frac{1}{1 + 2 \cos k} V_k (3x - 1),& 1/3\leq x <
      2/3,\\
      \frac{e^{\imath k} + 1}{1 + 2 \cos k} + 
      \frac{e^{-\imath k}}{1 + 2 \cos k} V^*_k(3x - 2),& 2/3\leq x <
      1,
    \end{array}
  \right.
  \label{snglebsgVk}
  \\
  W_k(x) = 
  \left\{
    \begin{array}{l@{\quad}l}
      \frac{e^{\imath k}}{1 + 2 \cos k} W_k (3 x),& 0\leq x < 1/3,\\
      \frac{e^{\imath k}}{1 + 2 \cos k} + 
      \frac{1}{1 + 2 \cos k} W^*_k (3x - 1),& 1/3\leq x <
      2/3,\\
      \frac{e^{\imath k} + 1}{1 + 2 \cos k} + 
      \frac{e^{-\imath k}}{1 + 2 \cos k} W_k(3x - 2),& 2/3\leq x <
      1,
    \end{array}
  \right.
  \label{snglebsgWk}
\end{eqnarray}
Equation \eref{snglebsgUk} defines the hydrodynamic modes of diffusion of
the triadic multi-baker map \cite{Gilbert:2001p356}. Equations
\eref{snglebsgVk} and \eref{snglebsgWk} correspond, on the other hand, to the
hydrodynamic modes of nine-adic multi-baker maps associated with random
walks with assigned probabilities $1/9$ to jump by two units to the left or
right, $2/9$ by a single unit, and $1/3$ to remain put. Notice in
particular that equation \eref{snglebsgVk} produces the  Sierpinski gasket
\cite{sierpinski:1915} for $k = \pi/3$, as shown in \fref{fig.sierp}. 

Further examples of piecewise linear maps and their associated eigenstates
can be found in \cite{driebe:1999}.

\section{\label{sec.conc}Concluding remarks}

In \cite{Tasaki:1995p226}, Tasaki and
Gaspard showed that the Takagi function describes the non-equilibrium
stationary state of a multi-baker map, and used its properties to show that
this system obeys Fick's law, which provides an expression of the mass
current as minus the product of the mass density gradient and diffusion
coefficient. This result thus provided a derivation of a phenomenological
law of thermodynamics in terms of the phase-space dynamics of the model,
which subsequently lead Gaspard to identify the fractality of the
non-equilibrium stationary states as the source of entropy production
\cite{Gaspard:1997p244}. A recent survey of these results with further
extensions can be found in \cite{Barra:2009p1004}. 

In this paper we have presented some of these results in the more 
general context of fractal curves and self-similar sets defined by
linear contractions, and emphasized their connections with the statistical
properties of maps that generate the associated functional equations. 
Shuichi Tasaki contributed in important ways to our understanding the
connection between singular functions and the statistical properties of
$r$-adic maps. We are deeply saddened that we are not able to talk to
Shuichi about the excursion into the world of fractals described here. We
know that we would have learned still more about them, and benefited from
his careful and clear explanations.

\ack{TG wishes to acknowledge financial support by the Belgian Federal
Government under the Inter-university Attraction Pole project NOSY
P06/02. He is affiliated with the Fonds de la Recherche
Scientifique F.R.S.-FNRS. JRD wishes to acknowledge the hospitality of the
Leviner Institute for Theoretical Physics at The Technion, Haifa, Israel.}

\end{document}